\documentclass[twocolumn]{aastex63}

\usepackage{amsmath}
\usepackage{empheq}

\accepted{by ApJ}
\shortauthors{LAPI \& DANESE}
\shorttitle{Stochastic Theory of the Hierarchical Clustering II.}

\newcommand*\widefbox[1]{\fbox{\hspace{2em}#1\hspace{2em}}}
\newcommand*\widewidefbox[1]{\fbox{\hspace{2em}#1\hspace{10.5em}}}

\begin{document}

\title{A Stochastic Theory of the Hierarchical Clustering\\
\small{II. Halo progenitor mass function and large-scale bias}}

\author[0000-0002-4882-1735]{Andrea Lapi}
\affiliation{SISSA, Via Bonomea 265, 34136 Trieste, Italy}\affiliation{IFPU - Institute for fundamental physics of the Universe, Via Beirut 2, 34014 Trieste, Italy}\affiliation{INFN-Sezione di Trieste, via Valerio 2, 34127 Trieste,  Italy}\affiliation{INAF-Osservatorio Astronomico di Trieste, via Tiepolo 11, 34131 Trieste, Italy}

\author{Luigi Danese}\affiliation{SISSA, Via Bonomea 265, 34136 Trieste, Italy}\affiliation{IFPU - Institute for fundamental physics of the Universe, Via Beirut 2, 34014 Trieste, Italy}

\begin{abstract}
We generalize the stochastic theory of hierarchical clustering presented in paper I by Lapi \& Danese (2020) to derive the (conditional) halo progenitor mass function and the related large-scale bias. Specifically, we present a stochastic differential equation that describes fluctuations in the mass growth of progenitor halos of given descendant mass and redshift, as driven by a multiplicative Gaussian white noise involving the power spectrum and the spherical collapse threshold of density perturbations. We demonstrate that, as cosmic time passes, the noise yields an average drift of the progenitors toward larger masses, that quantitatively renders the expectation from the standard extended Press \& Schechter (EPS) theory. We solve the Fokker-Planck equation associated to the stochastic dynamics, and obtain as an exact, stationary solution the EPS progenitor mass function. Then we introduce a modification of the stochastic equation in terms of a mass-dependent collapse threshold modulating the noise, and solve analytically the associated Fokker-Planck equation for the progenitor mass function. The latter is found to be in excellent agreement with the outcomes of $N-$body simulations; even more remarkably, this is achieved with the same shape of the collapse threshold used in paper I to reproduce the halo mass function. Finally, we exploit the above results to compute the large-scale halo bias, and find it in pleasing agreement with the $N-$body outcomes. All in all, the present paper illustrates that the stochastic theory of hierarchical clustering introduced in paper I can describe effectively not only halos' abundance, but also their progenitor distribution and their correlation with the large-scale environment across cosmic times.
\end{abstract}


\section{Introduction}\label{sec|intro}

In the standard picture of structure formation, dark matter halos grow hierarchically, progressively evolving toward larger and more massive units via mergers and accretion (see, e.g., textbooks by Mo et al. 2010; Cimatti et al. 2020). The leading-order halo statistics is constituted by the halo mass function $N(M,t)$, which provides the abundance of halo masses at any given cosmic epoch. However, to better characterize the growth history of halos a higher order statistics is needed, namely the progenitor mass function $\mathcal{N}(M_1,z_1|M_2,z_2)$; the latter specifies, for a descendant halo mass $M_2$ at final redshift $z_2$, the conditional distribution of progenitor halo masses $M_1<M_2$ at any redshift $z_1>z_2$.

Such a quantity can be exploited for a variety of purposes, such as constructing numerical realization of the merging process known as merger trees (see Lacey \& Cole 1993; Somerville \& Kolatt 1999; Cole et al. 2000; Parkinson et al. 2008; Zhang et al. 2008; Jiang \& van den Bosch 2014), computing halo formation time distributions and formation rates (see Lacey \& Cole 1993; Kitayama \& Suto 1996; Percival \& Miller 1999; Percival et al. 2000; Miller et al. 2006; Neistein \& Dekel 2008; Zhang et al. 2008; Moreno et al. 2009; Lapi et al. 2013), deriving the average growth history of halos (see Nusser \& Sheth 1999; van den Bosch 2002; Miller et al. 2006; Neistein et al. 2006), and investigating the relation of halos with their large-scale environment (see Mo \& White 1996; Sheth \& Tormen 1999; Musso et al. 2012; Paranjape et al. 2012, 2018).

Clearly, the halo progenitor distribution can be directly extracted from $N$-body simulations (see Kauffmann et al. 1999; Benson et al. 2000; Wechsler et al. 2002; Helly et al. 2003; Kang et al. 2005; Springel et al. 2005; McBride et al. 2009; Behroozi et al. 2013, 2019). However, this procedure is demanding in terms of time and storing capacity since it requires to run a full simulation and analyse its outputs; moreover, it comes together with some tricky and systematic issues, concerning not only the limited mass and force resolution, but also the algorithms used to identify and relate halos among different simulation snapshots (see, e.g., Harker et al. 2006; Fakhouri \& Ma 2008; Fakhouri et al. 2010; Jiang \& van den Bosch 2014). Thus a first principles theoretical understanding of the halo progenitor mass function is also very important.

The standard theoretical framework to address such an issue is constituted by the excursion set formalism and can be traced back to the seminal work by Bond et al. (1991) and to the ensuing developments by Lacey \& Cole (1993) and Mo \& White (1996). In this approach the issue of counting halos is remapped in finding the first crossing distribution of a random walk that hits a suitable barrier. Specifically, the random walk is executed by the overdensity field $\delta$ around a given spatial location as a function of the smoothed mass scale $\sigma^2(M)$, which encapsulates the power spectrum of perturbations; when the smoothing is performed with a $k-$sharp filter in Fourier space, the problem greatly simplifies since the random walk is Markovian. In addition, the barrier is provided by the linear collapse threshold $\delta_c(t,\sigma)$, which depends on cosmic time and possibly on the mass scale (in the latter case one deals with a so called ‘moving barrier'); when the mass-independent threshold $\delta_c(t)$ from the spherical collapse theory is used, the resulting framework is also known as extended Press \& Schechter theory (EPS)\footnote{Some authors in the literature tend to use `EPS' even in the general case of a moving barrier; throughout this paper we use `excursion set' for the general framework, and refer to `EPS' only in the special case of the mass-independent, spherical collapse threshold.}. The excursion set framework can be exploited not only to derive the halo mass function, but also other higher-order statistics like the progenitor distribution and bias; these quantities are related to the first crossing distributions for walks that do not start from the origin, and cross the barrier farer away from it on a larger mass scale.

Remarkably, the ensuing progenitor mass distribution is found to be in overall good agreement with the $N$-body outcomes, especially when a moving barrier with shape inspired by the ellipsoidal collapse is adopted (see Sheth \& Tormen 1999, 2002). A further bonus is that the progenitor distribution can be computed via closed-form analytic expressions (see Lacey \& Cole 1993; Sheth \& Tormen 2002; Mahmood \& Rajesh 2005) or via fast semi-analytic algorithms (see Zhang et al. 2008). This is useful both theoretically to transparently understand the physics underlying the hierarchial clustering of halos, and practically to investigate, much quickly than in simulations, the halo merger histories under diverse initial conditions and cosmological backgrounds.

Despite these successes and advantages, the excursion set framework is known to hide some pitfalls and drawbacks: the mere idea that one can relate the merging history of halos to the first upcrossing in the abstract $\delta_c-\sigma^2$ space constitutes a leap of faith (see Mo et al. 2010); the merging kernel associated to the excursion set theory is not symmetric, and this causes a mathematical inconsistency or at least an ambiguity in the definition of the merger rates (see Benson et al. 2005; Neistein \& Dekel 2008; Zhang et al. 2008); the relation between the probability of first upcrossing and the progenitor mass function cannot be strictly correct for individual mass elements (see discussion by Mo et al. 2010, their Sect. 7.2.2b).

In paper I (Lapi \& Danese 2020) we have submitted a new theory of the hierarchical clustering based on stochastic differential equations in real space, which constitutes a change of perspective with respect to the excursion set formalism and can alleviate the above shortcomings; our previous work has been focused on the (unconditional) halo mass function. In the present one we show how to generalize the treatment to the (conditional) progenitor mass distribution and large-scale halo bias.

The plan of the paper is the following. First, we present a stochastic differential equation that describes fluctuations in the mass growth of progenitor halos of given descendant mass and redshift, as driven by a multiplicative Gaussian white noise that involves the spherical collapse thresholds and the mass variances of the progenitor and descendant halos. In Sect.~\ref{sec|stochastic} we demonstrate that, as cosmic time passes, the noise yields an average drift of the progenitors toward larger masses, that quantitatively renders the expectation from the EPS theory. Then, in Sect.~\ref{sec|fokker}  we solve the Fokker-Planck equation associated to the stochastic dynamics, and obtain as a solution the EPS progenitor mass function;  we also point out that the solution is stationary when the original equation is written in terms of a convenient scaled variable. In Sect.~\ref{sec|driftdiff} we highlight how the formalism based on the Fokker-Planck equation allows to define unambiguously the total derivative of the (conditional and unconditional) mass function in terms of halo collapse vs. merging rates. In Sect.~\ref{sec|barrier} we introduce a minimal modification of the stochastic equation in terms of a mass-dependent collapse threshold modulating the noise, and solve analytically the associated Fokker-Planck equation for the progenitor mass function. The latter is found to be in excellent agreement with the outcomes of $N$-body simulations; even more remarkably, all that is achieved with the same mass-dependent threshold shape used in paper I to reproduce the unconditional halo mass function. In Sect.~\ref{sec|bias} we exploit the progenitor mass function to compute the large-scale halo bias, and show that it reproduces very well the simulation results. Finally, in Sect.~\ref{sec|summary} we summarize and briefly discuss our findings. In Appendix A, for the reader convenience, we recap and discuss the main assumptions and results of paper I, and we present an additional test concerning the unconditional mass function in warm DM cosmologies.

Unless otherwise explicitly specified, throughout this work we adopt the standard flat $\Lambda$CDM cosmology (Planck Collaboration 2020) with rounded parameter values: matter density $\Omega_M\approx 0.3$, dark energy density $\Omega_\Lambda\approx 0.7$, baryon density $\Omega_{\rm b}\approx 0.05$, Hubble constant $H_0 = 100\,h$ km s$^{-1}$ Mpc$^{-1}$  with $h\approx 0.7$, and mass variance $\sigma_8\approx 0.8$ on a scale of $8\, h^{-1}$ Mpc. The
most relevant equations are highlighted with a box.

\section{A stochastic equation for the EPS conditional probability}\label{sec|stochastic}

In analogy with the proposal put forward in paper I (see Appendix A for a summary), we now write down a nonlinear stochastic differential equation aimed to originate the conditional probability of the extended Press \& Schechter (EPS) theory. This reads (overdot means differentiation with respect to $t_1$)
\begin{empheq}[box=\fbox]{align}\label{eq|basic_sigma}
\cfrac{\rm d}{{\rm d}t_1}\,\ln (\Delta\sigma^2)^{-1/2} = \cfrac{(\Delta\sigma^2)^{1/2}}{\Delta\delta_c}\, \left|\cfrac{\Delta\dot{\delta_c}}{\Delta\delta_c}\right|^{1/2}\,\eta(t_1)~,
\end{empheq}
or equivalently, in terms of mass\footnote{Throughout the paper we adopt the Stratonovich convention, that allows to use the rules of ordinary calculus on stochastic variables, at the price of originating a noise-induced drift term in the Fokker-Planck coefficients associated to a given stochastic differential equation (see also paper I and Appendix A for details).}
\begin{equation}\label{eq|basic_mass}
\cfrac{\rm d}{{\rm d}t_1}\, M_1 = \left|\cfrac{\Delta\sigma^2}{\Delta\delta_c}\right|^{3/2}\,
\cfrac{2\,|\Delta\dot\delta_c|^{1/2}}{|{\rm d}\sigma_1^2/{\rm d}M_1|}\; \eta(t_1)~.
\end{equation}
Here $t$ is the cosmic time and $\eta(t)$ is a Gaussian white noise (physical dimension $1/\sqrt{\rm time}$) with ensemble-average properties $\langle\eta(t)\rangle=0$ and $\langle\eta(t)\, \eta(t')\rangle=2\, \delta_{\rm D}(t-t')$, where $\delta_{\rm D}$ is the Dirac $\delta-$function.
In the above the subscript $2$ refers to the quantities at the final cosmic time $t_2$ of the descendant halo, and the subscript $1$ refers to those at the cosmic time $t_1$ of the progenitor halos; keeping this in mind, we have defined
$\Delta\delta_c\equiv \delta_c(t_1)-\delta_c(t_2)$ and $\Delta\sigma^2\equiv \sigma^2(M_1)-\sigma^2(M_2)=\sigma_1^2-\sigma_2^2$. The quantity $\delta_c(z)=\delta_{c0}\,D(0)/D(z)$ is the critical threshold for collapse extrapolated from linear perturbation theory; in a flat Universe (see Mo et al. 2010; Weinberg 2008), one can use the approximations $\delta_{c0} \simeq \frac{3}{20}\, (12\pi)^{2/3}$ $\left[1+0.0123\log_{10}\Omega_M(z)\right]\approx 1.68$
and $D(z) \approx \frac{5}{2}\,\frac{\Omega_{M}(z)}{1+z}\,\left[\frac{1}{70}+
\frac{209}{140}{\Omega_M(z)-\frac{1}{140}\,\Omega_M^2(z)+\Omega_M^{4/7}(z)}\right]^{-1}$ with $\Omega_M(z)\equiv \Omega_M\,(1+z)^3/[\Omega_\Lambda+\Omega_M\,(1+z)^3]$. The quantity $\sigma$ is the mass variance filtered on the mass scale $M$:
\begin{equation}\label{eq|variance}
\sigma^2(M) = \cfrac{1}{(2\pi)^3}\,\int{\rm d}^3k\, P(k)\,\tilde{W}_M^2(k)~,
\end{equation}
in terms of the power spectrum $P(k)$ and of a Fourier transformed window function $\tilde{W}_M^2(k)$ whose volume in real space encloses the mass $M$;
$\sigma(M)$ is a deterministic function of $M$, that
for standard cold dark matter power spectra (e.g., Bardeen et al. 1986) features an inverse, convex, slowly-varying behavior. In the present framework $M(t)$ and $\sigma(t)=\sigma(M(t))$ constitute stochastic variables
that fluctuate over cosmic time $t$ under the influence of the noise.
A discussion of the key assumptions underlying Eq.~(\ref{eq|basic_sigma}) and a summary of the ensuing stochastic framework is presented in paper I and summarized in Appendix A.
%According to the above In the present theory the relation $\sigma(M)$ between $\sigma$ and $M$ is purely deterministic, but both $M(t)$ and $\sigma(M(t))$ are to be considered stochastic variables that
%; assuming different filter functions (e.g., Gaussian or top-hat in real or in Fourier space) changes only the deterministic relation $\sigma(M)$ but has otherwise no effect on the Markovianity of the stochastic processes $M(t)$ and $\sigma(M(t))$.

\begin{figure}
\centering
\includegraphics[width=\textwidth]{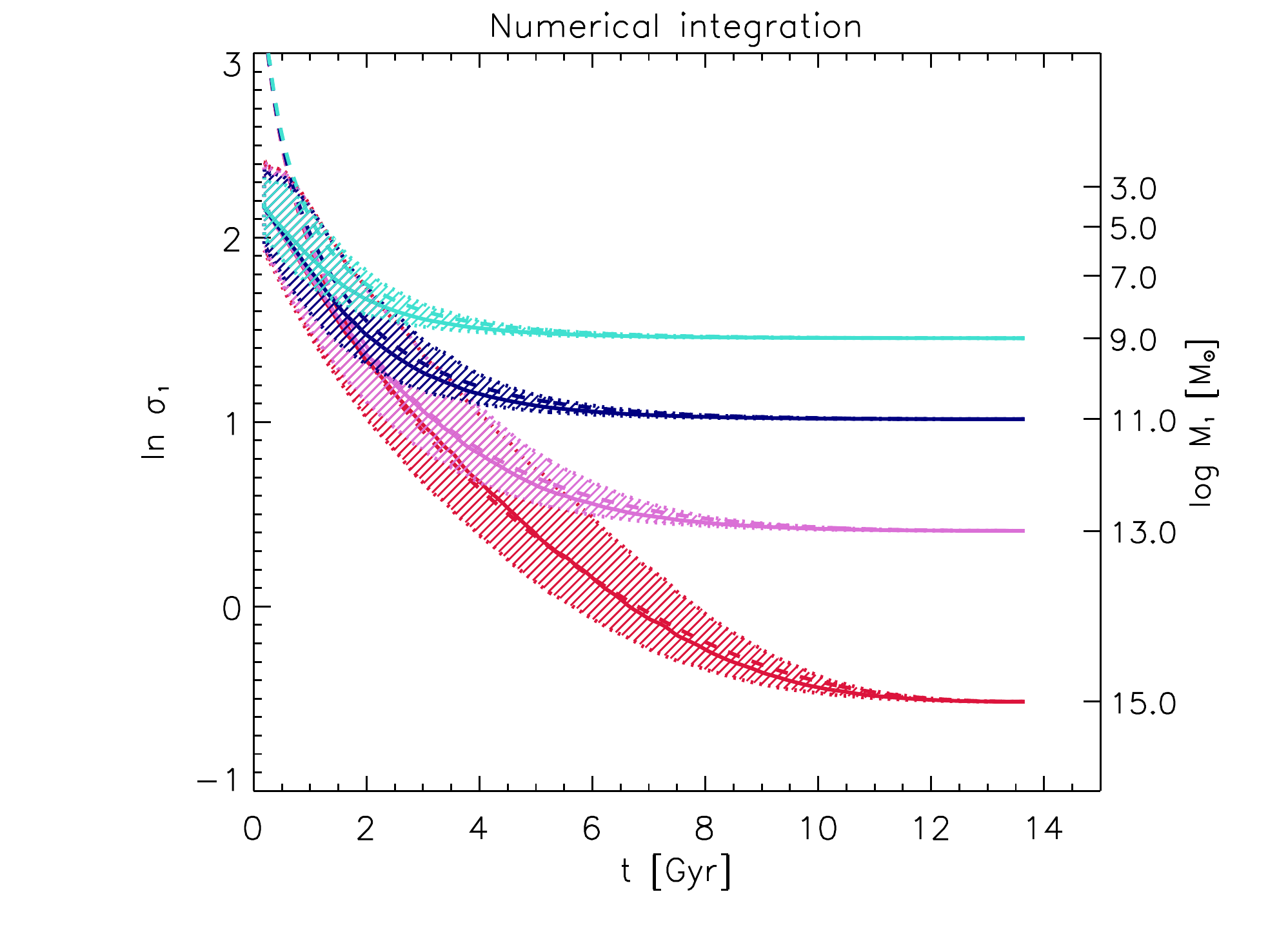}
\caption{Numerical integration of the stochastic differential Eq.~(\ref{eq|basic_sigma}), yielding the evolution of the progenitors' mass variance $\ln\sigma_1$ (left $y-$axis) and of the progenitors' mass $M_1$ (right $y-$axis) as a function of cosmic time $t$; the initial condition $M_1\approx 10^4\, M_\odot$ at $z\sim 10$ (corresponding to $t_1\sim 0.5$ Gyr) has been adopted. Solid lines surrounded by shaded areas illustrate the median and the quartiles over $3000$ realization of the noise; different colors are for different final masses $M_2\approx 10^9$ (cyan), $10^{11}$ (blue), $10^{13}$ (magenta) and $10^{15}\, M_\odot$ (red) at $z_2\approx 0$. The dashed lines show the evolution for a constant $\nu_{12}\equiv \Delta\delta_c/\sqrt{\Delta\sigma^2}$, as expected in the EPS theory.}\label{fig|sims_cond}
\end{figure}

Assigned the descendant halo mass $M_2$ at cosmic time $t_2$ and an initial condition at early times, the above equation can be integrated via the Euler-Maruyama or Euler-Heun methods (e.g. Kloeden \& Platen 1992) on a discretized time grid. % $t_{1,k}$ with $k=0,\ldots,N-1$ such that $t_{1,N}\equiv t_2$; the solution is
%\begin{equation}\label{eq|Euler}
%\ln\sigma(t_{1,i+1})=\ln\sigma(t_{1,i})-\cfrac{1}{\sigma^2(t_{1,i})}\,
%\left|\cfrac{\sigma^2(t_{1,i})-\sigma^2(t_2)}{\delta_c(t_{1,i})-\delta_c(t_2)}\right|^{3/2}\, \left|\delta_c(t_{1,i+1})-\delta_c(t_{1,i})\right|^{1/2}\, w_i~,
%\end{equation}
%where $w_i$ are random weights extracted from a normal distribution with zero mean and unit variance.
In Fig.~\ref{fig|sims_cond} we show the resulting evolution of $\ln\sigma(M_1(t))$ and of $\log M_1(t)$ as a function of cosmic time, with initial condition $M_1\sim 10^4\, M_\odot$ or $\ln\sigma_1\sim 2.2$ at $z\sim 10$ (reasonable initial conditions do not change significantly the outcome). The colored solid lines surrounded by shaded areas illustrate the median and the quartiles over $3000$ realization of the noise, for different
descendant masses $M_2\approx 10^9-10^{11}-10^{13}-10^{15}\, M_\odot$ at redshift $z_2\approx 0$; the corresponding colored dashed lines show the expected evolution when the quantity $\nu_{12}\equiv \Delta\delta_c/\sqrt{\Delta\sigma^2}$ stays constant, as expected in the EPS framework. It is found that the noise induces a drift of $M_1$ ($\sigma_1$), making it to increase (decrease) and, after a burn-in period needed to erase memory of the initial condition, to converge toward the evolution predicted by the EPS theory.

\subsection{Fokker-Planck equation and the EPS mass function}\label{sec|fokker}

The probability density $\mathcal{P}(M_1,t_1|M_2,t_2)$ of finding a progenitor halo with mass between $M_1$ and $M_1+{\rm d}M_1$ at cosmic time $t_1$ given a descendant halo with mass $M_2$ at cosmic time $t_2$, can be derived by solving the Fokker-Planck equation associated to Eq.~(\ref{eq|basic_mass}). This reads (see Appendix A in paper I for details; also textbooks by Risken 1996 and Paul \& Baschnagel 2013)
\begin{equation}\label{eq|fokker}
\partial_{t_1} \mathcal{P} = -\mathcal{T}^2\, \partial_{M_1}\,\left[\mathcal{D}\, (\partial_{M_1}\mathcal{D})\,\mathcal{P}\right] +\mathcal{T}^2\,\partial_{M_1}^2\,\left[\mathcal{D}^2\, \mathcal{P}\right]~,
\end{equation}
where we have defined the two quantities $\mathcal{D}(M_1,M_2)\equiv 2\,(\Delta\sigma^2)^{3/2}/|{\rm d}\sigma^2/{\rm d}M_1|$ and $\mathcal{T}(t_1,t_2)\equiv |\dot \Delta\delta_c|^{1/2}/(\Delta\delta_c)^{3/2}$ such that in Eq.~(\ref{eq|basic_mass}) the mass and time dependencies are factorized as $\dot M_1 = \mathcal{D}(M_1,M_2)\, \mathcal{T}(t_1,t_2)\, \eta(t_1)$. The Fokker-Planck equation may also be written as a pure continuity equation $\partial_{t_1} \mathcal{P}+\partial_{M_1}\, \mathcal{J}=0$ in terms of a probability current $\mathcal{J}=-\mathcal{T}^2\,\mathcal{D}\, \partial_{M_1}\, [\mathcal{D}\, \mathcal{P}]$. The natural boundary conditions $\lim_{M_1\rightarrow \infty} \mathcal{P}=0$, $\mathcal{P}(M_1,0|M_2,t_2)=\delta_{\rm D}(M_1)$ and the constraint $\mathcal{P}(M_1,t_1|M_2,t_2)=0$ whenever $M_1<0$ or $M_1>M_2$ must apply; the latter correspond to reflective barrier conditions $\mathcal{J}|_{M_1=0,M_1=M_2}=[\mathcal{D}\, \partial_{M_1}\,(\mathcal{D}\,\mathcal{P})]|_{M_1=0,M_1=M_2}=0$ at the $M_1=0$ and $M_1=M_2$ points (no net probability currents through $M_1=0$ or $M_1=M_2$).
Now to solve the Fokker-Planck equation we employ the transformations:
\begin{equation}\label{eq|auxvar}
\left\{
\begin{aligned}
& X\equiv \int\cfrac{{\rm d}M_1}{\mathcal{D}(M_1,M_2)} = \cfrac{1}{(\Delta\sigma^2)^{1/2}}\\
\\
& Y\equiv \int{\rm d}t_1\, \mathcal{T}^2(t_1,t_2)=\cfrac{1}{2\, (\Delta\delta_c)^2}\\
\\
& \mathcal{W}\equiv \mathcal{D}\, \mathcal{P}~.
\end{aligned}
\right.
\end{equation}
Then Eq.~(\ref{eq|fokker}) turns into
\begin{equation}\label{eq|diffeq}
\partial_Y\, \mathcal{W} = \partial_X^2\, \mathcal{W}~,
\end{equation}
which is a standard diffusion equation with boundary conditions $\lim_{X\rightarrow \infty}\mathcal{W}=0$, $\mathcal{W}(X,0)=\delta_{\rm D}(X)$ and $(\partial_X\,\mathcal{W})|_{X=0,X=\infty}=0$, whose solution just writes $\mathcal{W} = (\pi\, Y)^{-1/2}\, e^{-X^2/4\, Y}$.
Coming back to the original variables we get the EPS conditional probability
\begin{widetext}
\begin{empheq}[box=\widefbox]{align}\label{eq|fokkersol}
\mathcal{P}(M_1,t_1|M_2,t_2) = \cfrac{1}{\sqrt{2\pi}}\, \cfrac{\Delta\delta_c}{(\Delta\sigma^2)^{3/2}}\,\left|\cfrac{{\rm d}\sigma^2}{{\rm d}M_1}\right|\, e^{-(\Delta\delta_c)^2/2\,\Delta\sigma^2}~.
\end{empheq}
\end{widetext}
Note that in the limit $\delta_c(t_2)\rightarrow 0$ and $\sigma(M_2)\rightarrow 0$ the above probability converges to
\begin{equation}\label{eq|PSlimit}
\mathcal{P}(M,t) = \sqrt{\cfrac{2}{\pi}}\, \cfrac{\delta_c}{\sigma^2}\,\left|\cfrac{{\rm d}\sigma}{{\rm d}M}\right|\, e^{-(\delta_c)^2/2\,\sigma^2}~,
\end{equation}
which is related to the unconditional mass function by $N(M,t)=(\bar\rho/M)\, \mathcal{P}(M,t)$ in terms of the comoving matter density $\bar\rho$; as pointed out in paper I, this is exactly the Press \& Schechter formula.

We can derive the same results in another way, just by recasting Eq.~(\ref{eq|basic_sigma}) in terms of the scaled (or self-similar) variable $\nu_{12}\equiv \Delta\delta_c/\sqrt{\Delta\sigma^2}$ as
\begin{equation}\label{eq|basic_nu}
\cfrac{\rm d}{{\rm d}t_1}\,\nu_{12} = -\left|\cfrac{\Delta\dot\delta_c}{\Delta\delta_c}\right|\,\nu_{12}+ \left|\cfrac{\Delta\dot\delta_c}{\Delta\delta_c}\right|^{1/2}\, \eta(t_1)~.
\end{equation}
The corresponding Fokker-Planck equation for the probability $\mathcal{P}(\nu_{12},t)$ reads
\begin{equation}\label{eq|fokkernu}
\partial_{t_1}\,\mathcal{P}=\left|\cfrac{\Delta\dot\delta_c}{\Delta\delta_c}\right|\,
\partial_{\nu_{12}}\,\left[\nu_{12}\,\mathcal{P}+\partial_{\nu_{12}}\,\mathcal{P}\right]~,
\end{equation}
with boundary conditions $\lim_{\nu_{12}\rightarrow \infty}\mathcal{P}=0$ and $\mathcal{J}|_{\nu_{12}=0}=-|\dot\Delta\delta_c/\Delta\delta_c|\,[\nu_{12}\,\mathcal{P}+\partial_{\nu_{12}}\,\mathcal{P}]|_{\nu_{12}=0}=0$,
implying $\int_0^{\infty}{\rm d}\nu_{12}\,\mathcal{P}=1$.

Under the ergodic hypothesis, we can focus on the stationary solution $\mathcal{P}(\nu_{12},t_1)=\bar{\mathcal{P}}(\nu_{12})$; this is determined by $\partial_{t_1}\,\mathcal{P}(\nu_{12},t_1)=0$ or
\begin{equation}
\nu_{12}\,\bar{\mathcal{P}}+\cfrac{{\rm d}}{{\rm d}\nu_{12}}\,\bar{\mathcal{P}}=0~,
\end{equation}
where the constant on the r.h.s. must be zero to satisfy the no-current boundary condition $\mathcal{J}|_{\nu_{12}=0}=0$. The ensuing solution is just $\bar{\mathcal{P}}(\nu_{12})=\sqrt{2/\pi}\, e^{-\nu_{12}^2/2}$. The conditional probability function reads
\begin{equation}\label{eq|massfuncnu}
\mathcal{P}(M_1,t_1|M_2,t_2) = \left|\cfrac{{\rm d}\nu_{12}}{{\rm d}M_1}\right|\, \bar{\mathcal{P}}(\nu_{12})~;
\end{equation}
since $|{\rm d}\nu_{12}/{\rm d}M_1| = \Delta\delta_c/2\,(\Delta\sigma^2)^{3/2} \times |{\rm d}\sigma_1^2/{\rm d}M_1|$, this again boils down to the EPS probability of Eq.~(\ref{eq|fokkersol}).

Note that often in the literature the quantity $\nu_{12}\, \bar{\mathcal{P}}(\nu_{12})$ is referred to as conditional multiplicity function, while the quantity $f(\log M_1|M_2)\equiv \mathcal{P}(M_1,t_1|M_2,t_2)\,M_1\,\ln(10)$
is the mass fraction in progenitor haloes per logarithmic mass bin. Finally, the (conditional) progenitor mass function $\mathcal{N}(M_1,t_1|M_2,t_2)$, which represents the ensemble-averaged number of progenitors with mass in the range $M_1$ and $M_1+{\rm d}M_1$ at time $t_1$ for a halo of mass $M_2$ at current time $t_2$ is given by
\begin{widetext}
\begin{equation}\label{eq|prog_mf}
\mathcal{N}(M_1,t_1|M_2,t_2) = \cfrac{M_2}{M_1}\, \mathcal{P}(M_1,t_1|M_2,t_2) = \cfrac{M_2}{M_1}\,\left|\cfrac{{\rm d}\nu_{12}}{{\rm d}M_1}\right|\, \bar{\mathcal{P}}(\nu_{12})~.
\end{equation}
\end{widetext}

\begin{figure}
\centering
\includegraphics[width=0.9\textwidth]{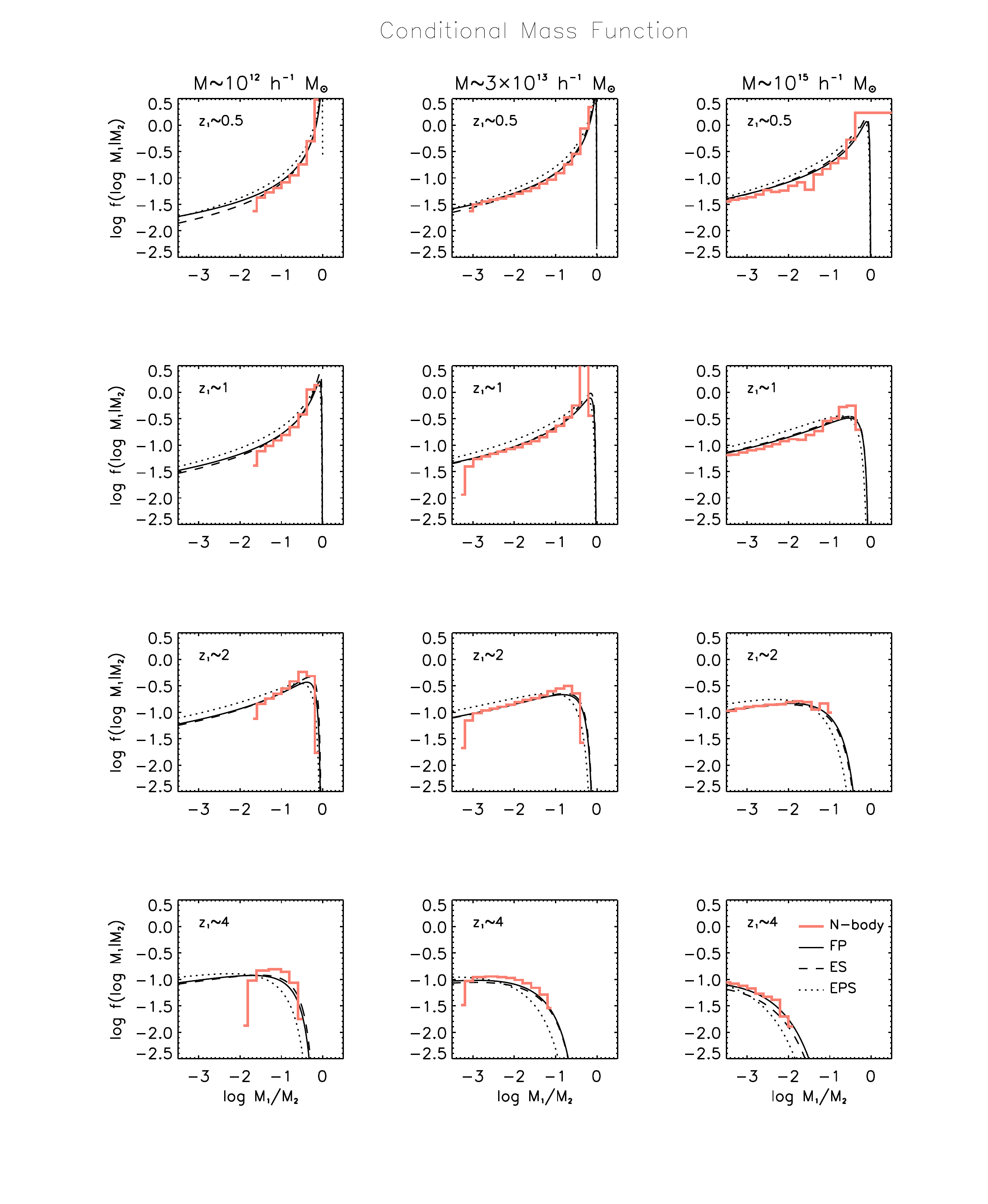}
\caption{The mass fraction in progenitor haloes $f(\log M_1|M_2)\equiv \mathcal{P}(M_1,t_1|M_2,t_2)\, M_1\,\ln(10)$ as a function of the progenitor to descendant mass ratio $M_1/M_2$; panels are for different descendant masses $M_2$ at $z_2\approx 0$ (in columns) and for different progenitor redshifts $z_1$ (in rows), as labeled. The orange histograms illustrate the outcomes of the \emph{Millennium} $N$-body simulations for FoF-identified halos by Cole et al. (2008). The solid line reports our stochastic theory based on the Fokker-Planck solution of Eq.~(\ref{eq|fokkersol_barrier_full}) for a mass-dependent threshold $\Delta\delta_c(\sigma,t)$ modulating the noise, as given by Eqs.~(\ref{eq|barrier}) and (\ref{eq|barrauxvar}); the same shape parameters used in paper I to reproduce the unconditional halo mass function have been retained. Dashed line shows the corresponding result for the excursion set framework with a moving barrier. Dotted line refers to the EPS theory, i.e. a standard spherical collapse threshold $\Delta\delta_c(t)$ independent of mass. All the analytic results have been computed with the same cosmological parameters of the \emph{Millennium} $N$-body simulations for fair comparison.
}\label{fig|condmassfunc}
\end{figure}

In Figs.~\ref{fig|condmultfunc} and \ref{fig|condmassfunc} we illustrate as dotted lines the conditional multiplicity function and the mass fraction in progenitors derived above. As it is well known from EPS theory, such conditional mass functions tend to exceed that the outcomes of $N$-body simulations at the low-mass end, and to fall short of them at the high mass end.

\subsection{Drift vs. diffusion}\label{sec|driftdiff}

Before proceeding further toward a better agreement with simulations, it is interesting to highlight the relative role played by the two terms appearing on the right hand side of the Fokker-Planck equation in determining the total derivative of the (conditional and unconditional) mass function. To this purpose, after Eqs.~(\ref{eq|fokker}) and (\ref{eq|prog_mf}) one can write
\begin{equation}\label{eq|driftdiff}
\left\{
\begin{aligned}
& \partial_{t_1}\, \mathcal{N} = (\partial_{t_1}\, \mathcal{N})_{\rm DRIFT}+(\partial_{t_1}\, \mathcal{N})_{\rm DIFF}~, \\
\\
& (\partial_{t_1}\, \mathcal{N})_{\rm DRIFT} = -\cfrac{\mathcal{T}^2}{M_1}\, \partial_{M_1}\,\left[\mathcal{D}\, (\partial_{M_1}\mathcal{D})\,M_1\,\mathcal{N}\right]~,\\
\\
&(\partial_{t_1}\, \mathcal{N})_{\rm DIFF} =+\cfrac{\mathcal{T}^2}{M_1}\,\partial_{M_1}^2\,\left[\mathcal{D}^2\, M_1\,\mathcal{N}\right]~.
\end{aligned}
\right.
\end{equation}
where the quantities $\mathcal{D}$ and $\mathcal{T}$ have been defined below Eq.~(\ref{eq|fokker}). Here the term $(\partial_{t_1}\, \mathcal{N})_{\rm DRIFT}$ stands for drift, and can be interpreted as the direct hierarchical collapse of larger and more massive perturbations; the term $(\partial_{t_1}\, \mathcal{N})_{\rm DIFF}$ stands for diffusion, and can be interpreted as describing the effect of stochastic merging events.

\begin{figure}
\centering
\includegraphics[width=\textwidth]{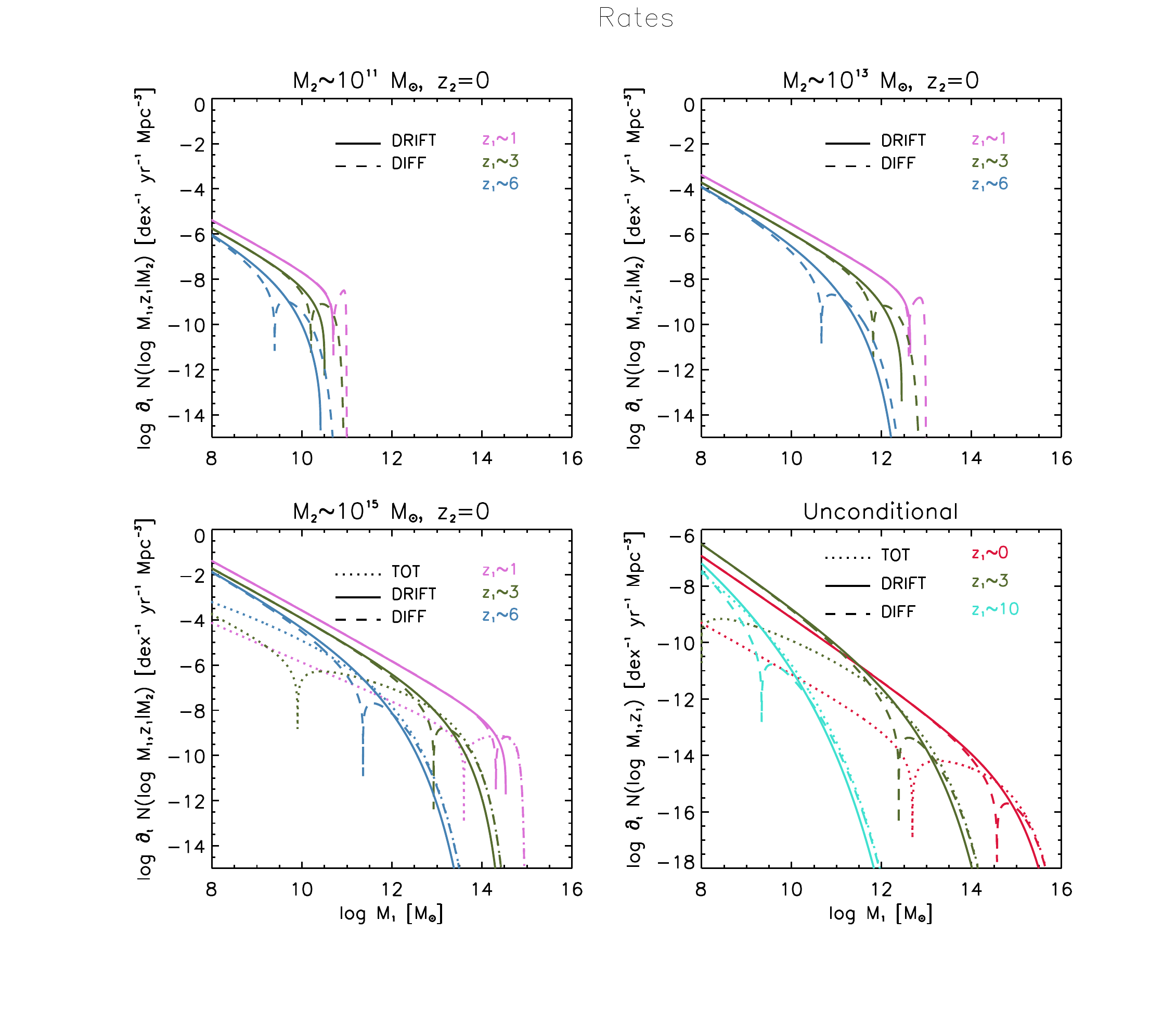}
\caption{Time derivative (logarithm of the absolute value) of the conditional mass function (spherical collapse) as a function of the progenitor mass $M_1$. Top left panel shows the conditional rate for descendant mass $M_2\approx 10^{11}\, M_\odot$ and redshift $z_2\approx 0$, top right shows the same quantity for $M_2\approx 10^{13}\, M_\odot$ and $z_2\approx 0$, bottom left shows the same for $M_2\approx 10^{15}\, M_\odot$ and $z_2\approx 0$, and bottom right panel shows the unconditional rate. In each panel solid lines refer to the drift contribution, dashed lines to the diffusion contribution and dotted lines (only shown in bottom panels for clarity) to the total time derivative. Color-code refers to different redshifts $z_1\approx 0$ (red), $1$ (magenta), $3$ (green), $6$ (blue) and $10$ (cyan), as labeled in each panel. }\label{fig|rates}
\end{figure}

In Fig.~\ref{fig|rates} we illustrate the contributions (in absolute value) of the drift (solid lines) and diffusion (dashed line) terms to the time derivative of the conditional mass function $\partial_{t_1}\mathcal{N}$; these are plotted as a function of the progenitor mass $M_1$ at various progenitor redshifts $z_1$ (color-coded), for different descendant masses $M_2\approx 10^{11}$ (top left panel), $10^{13}$ (top right panel) and $10^{15}\, M_\odot$ (bottom left panel) at $z_2\approx 0$. The bottom right panel shows the same for the unconditional mass function.

The drift term is always positive (despite the minus sign appearing in Eq.~\ref{eq|driftdiff}), while the diffusion term is positive at the high mass end, and becomes negative toward lower masses. The overall effect of the former is to cause an upward drift of the mass function toward larger masses, while that of the latter is to reshape the mass function by subtracting halos from the low-mass end and incorporating them into more massive objects;  this is indeed what is to be expected from the combined action of direct collapse and stochastic mergers.

All in all, the total time derivative of the (conditional and unconditional) mass function is positive and dominated by diffusion at the high mass end (new halos originate from merging of smaller ones), it is positive and dominated by drift at intermediate masses (new halos collapse from the perturbation field), while it becomes negative and is again dominated by diffusion toward the low mass end (small masses are destroyed by merging into larger ones). Finally, the impact of diffusion is more relevant toward relatively larger masses and later cosmic times, while the drift term becomes more and more important over a progressively extended mass range at higher redshifts.

We stress that the separation of the total time derivative of the mass function in a drift vs. a diffusion term, and the related physical interpretation in terms of direct collapse vs. merging, is exclusively allowed by our stochastic theory based on the Fokker-Planck equation. Similar arguments in the standard EPS framework prove to be difficul (e.g., Benson et al. 2005; Neistein \& Dekel 2008; Zhang et al. 2008; Lapi et al. 2013), and actually originate tricky ambiguities in the definition of the halo collapse and merging rates. An interesting future development, which goes beyond the scope of the present paper, would be to construct refined merger trees based on the above Fokker-Planck separation between the direct collapse and merging rates; speculating somewhat, this could help alleviating some tensions between the canonic merger-tree based approaches to galaxy formation and high-redshift observations (especially related to massive, strongly star-forming systems).

\section{Mass-dependent threshold: the $N$-body progenitor mass function}\label{sec|barrier}

In paper I we have shown that the $N$-body unconditional mass function can be reproduced in our stochastic approach by including in the multiplicative noise term a mass dependent collapse threshold with shape
\begin{equation}\label{eq|barrier}
\delta_c(\sigma,t) = \sigma\, \sqrt{q}\, \nu\, \left[1+\cfrac{\beta}{(\sqrt{q}\, \nu)^{2\,\gamma}}\right]~,
\end{equation}
in terms of three parameters $q$, $\beta$, and $\gamma$; comparison with the $N$-body outcomes (see Sheth \& Tormen 1999; Bhattacharya et al. 2011; Watson et al. 2013) indicate the approximate values $q\approx 0.65$, $\beta\approx 0.15$ and $\gamma\approx 0.35$. Note that in the excursion set approach, $\delta_c(\sigma,t)$ acts as a moving barrier, whose shape is inspired by the ellipsoidal collapse. In the present stochastic theory it is just a quantity modulating the noise, meant to describe a bunch of effects not only limited to the non-spherical shape of perturbations, but also to other complex phenomena that can influence their collapse, like tidal torques and angular momentum, dynamical friction, and possibly the speciality of collapse locations (see paper I and Appendix A for discussion).

Now we generalize the approach to derive the conditional probability function and show that, with the same shape of the collapse threshold and parameter values given above, it performs very well in reproducing the progenitor distribution extracted from $N$-body simulations. The basic idea is just to replace $\Delta\delta_c\equiv \delta_c(t_1)-\delta_c(t_2)$ with $\Delta B(\nu_{12})\equiv \delta_c(\sigma_1,t_1)-\delta_c(\sigma_2,t_2)$ in the multiplicative noise term of Eq.~(\ref{eq|basic_sigma}); when formulated in terms of the scaled variable $\nu_{12}\equiv \Delta \delta_c/\sqrt{\Delta\sigma^2}$, one gets
\begin{equation}\label{eq|basic_barrier}
\cfrac{\rm d}{{\rm d}t_1}\,\nu_{12} = -\nu_{12}\,\left|\cfrac{\Delta\dot\delta_c}{\Delta\delta_c}\right| +\cfrac{\Delta\delta_c}{\Delta B(\nu_{12})}\,\left|\cfrac{\Delta\dot\delta_c}{\Delta\delta_c}\right|^{1/2}\, \eta(t_1)~,
\end{equation}
where $\Delta B(\nu_{12})\equiv \Delta \delta_c\,[B_0+B_1\,(1+B_2/\nu_{12}^2)^\gamma]$ can be expressed in terms of the coefficients
\begin{equation}\label{eq|barrauxvar}
\left\{
\begin{aligned}
& B_0\equiv \sqrt{q}-\beta\,\cfrac{\sqrt{q}\,\delta_2}{\Delta\delta_c}\, \left(\cfrac{\sigma_2}{\sqrt{q}\,\delta_2}\right)^{2\,\gamma}\\
\\
& B_1\equiv \beta\,\cfrac{\sqrt{q}\,\delta_1}{\Delta\delta_c}\, \left(\cfrac{\sigma_2}{\sqrt{q}\,\delta_1}\right)^{2\,\gamma}\\
\\
& B_2\equiv \left(\cfrac{\Delta\delta_c}{\sigma_2}\right)^2~.
\end{aligned}
\right.
\end{equation}
Note that the spherical collapse threshold independent of mass is recovered in the limit $q\rightarrow 1$ and $\beta\rightarrow 0$, which readily imply $B_0\simeq 1$ and $B_1\simeq 0$; then $\Delta B(\nu_{12})\simeq \Delta\delta_c$ holds and Eq.~(\ref{eq|basic_barrier}) collapses into Eq.~(\ref{eq|basic_nu}).

The corresponding Fokker-Planck equation reads:
\begin{equation}\label{eq|fokker_barrier}
\partial_{t_1}\,\mathcal{P}=\left|\cfrac{\Delta\dot\delta_c}{\Delta\delta_c}\right|\,
\partial_{\nu_{12}}\,\left\{\nu_{12}\,\mathcal{P}+\cfrac{\Delta\delta_c}{\Delta B(\nu_{12})}\,\partial_{\nu_{12}}\,
\left[\cfrac{\Delta\delta_c}{\Delta B(\nu_{12})}\,\mathcal{P}\right]\right\}~,
\end{equation}
and the relevant stationary solution $\mathcal{P}=\bar{\mathcal{P}}(\nu_{12})$ is determined by
\begin{equation}
\cfrac{\Delta\delta_c}{\Delta B(\nu_{12})}\,\cfrac{{\rm d}}{{\rm d}\nu_{12}}\,
\left[\cfrac{\Delta \delta_c}{\Delta B(\nu_{12})}\,\bar{\mathcal{P}}\right] = -\nu_{12}\,\bar{\mathcal{P}}~,
\end{equation}
where the integration constant must be null to satisfy the no-current boundary condition $\mathcal{J}|_{\nu_{12}=0}=0$. The above equation can be easily solved by multiplying both sides by $\Delta\delta_c/\Delta B(\nu_{12})$ and recognizing that it becomes  separable for the function $[\Delta\delta_c/\Delta B(\nu_{12})]\,\bar{\mathcal{P}}$; we find
\begin{widetext}
\begin{equation}\label{eq|fokkersol_barrier}
\bar{\mathcal{P}}(\nu_{12})= \mathcal{A}_{12}\,\cfrac{\Delta B(\nu_{12})}{\Delta\delta_c}\,\exp\left\{-\int{\rm d}\nu_{12}\,\nu_{12}\,\left[\cfrac{\Delta B(\nu_{12})}{\Delta\delta_c}\right]^2\right\}~,
\end{equation}
\end{widetext}
where the normalization constant $\mathcal{A}_{12}$ is determined by the condition $\int_0^\infty{\rm d}\nu_{12}\, \bar{\mathcal{P}}(\nu_{12})=1$.
Inserting the full shape of $\Delta B(\nu_{12})$ and performing explicitly the integration yields the closed form expression
\begin{widetext}
\begin{empheq}[box=\widewidefbox]{align}\label{eq|fokkersol_barrier_full}
\begin{aligned}
\bar{\mathcal{P}}_{\rm CMF}(\nu_{12}) &= \mathcal{A}_{12}\,B_0\,\left[1+\cfrac{B_1}{B_0}\,\left(1+\cfrac{B_2}{\nu_{12}^2}\right)^\gamma\right]\,\exp\Bigg\{-\cfrac{B_0^2}{2}\,\nu_{12}^2\,\bigg[1+\\
\\
&+\left(\cfrac{B_1}{B_0}\right)^2\,\left(\cfrac{1+B_2/\nu_{12}^2}{1+\nu_{12}^2/B_2}\right)^{2\gamma}\,
\cfrac{{}_2F_1(1-2\gamma,-2\gamma,2-2\gamma;-\nu_{12}^2/B_2)}{1-2\gamma}+\\
\\ &+2\,\cfrac{B_1}{B_0}\,\left(\cfrac{1+B_2/\nu_{12}^2}{1+\nu_{12}^2/B_2}\right)^{\gamma}\,\cfrac{{}_2F_1(1-\gamma,-\gamma,2-\gamma;-\nu_{12}^2/B_2)}
{1-\gamma}\bigg]\Bigg\}~.
\end{aligned}
\end{empheq}
\end{widetext}
In the above ${}_2F_1$ is the ordinary hypergeometric function, which is defined by the power series ${}_2F_1(a,b,c;x)\equiv \sum_{n=0}^\infty (a)_n\,(b_n)\,x^n/(c)_n\,n!$ in terms of the Pochammer's symbols $(q)_0=1$ and $(q)_n=q\,(q+1)\,\ldots\,(q+n-1)$ for any integer $n>0$; note that in the limit $x\rightarrow 0$ one has ${}_2F_1(a,b,c;x)\simeq 1+a\,b\,x/c+\ldots$.

To recover the unconditional probability (linked to the halo mass function) from the above expression, it is sufficient to take the limits $\sigma_2\rightarrow 0$ and $\delta_2\rightarrow 0$. In such a case, $B_0\simeq \sqrt{q}$, $B_1\simeq 0$, $B_2\rightarrow \infty$ but $B_1\,B_2^\gamma\simeq \beta\,(\sqrt{q})^{1-2\gamma}$, to imply $\nu_{12}=\delta_1/\sigma_1\equiv \nu$ and
\begin{widetext}
\begin{equation}\label{eq|fokkersol_barrier_uncond}
\bar{\mathcal{P}}_{\rm MF}(\nu)\simeq \mathcal{A}\, \sqrt{q}\,\left[1+\cfrac{\beta}{(\sqrt{q}\,\nu)^{2\gamma}}\right]\, \exp\left\{-q\,\cfrac{\nu^2}{2}\,\left[1+\cfrac{2\,\beta}{1-\gamma}\,\cfrac{1}{(\sqrt{q}\,\nu)^{2\gamma}}
+\cfrac{\beta^2}{1-2\,\gamma}\,\cfrac{1}{(\sqrt{q}\,\nu)^{4\gamma}}\right]\right\}~,
\end{equation}
\end{widetext}
which is the result derived in paper I (see also Appendix A).

\begin{figure}
\centering
\includegraphics[width=\textwidth]{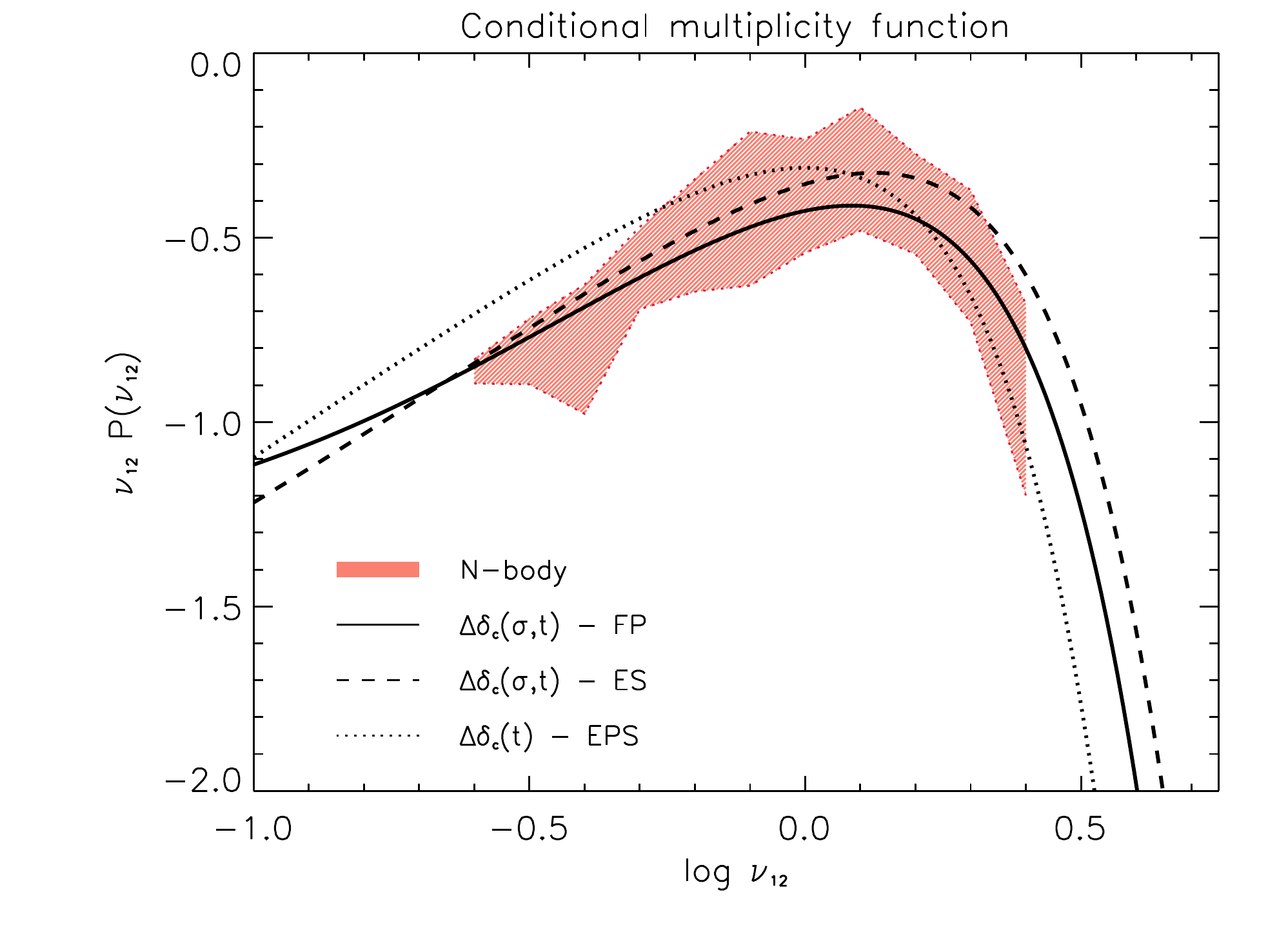}
\caption{The conditional multiplicity function $\nu_{12}\, \bar{\mathcal{P}}(\nu_{12})$ as a function of the self-similar, scaled variable $\nu_{12}\equiv \Delta\delta_c/\sqrt{\Delta\sigma^2}$, see Sect.~\ref{sec|stochastic} for details. The orange shaded area illustrates the outcome of the \emph{Millennium} $N$-body simulations for FoF-identified halos by Cole et al. (2008). The solid line reports our stochastic theory based on the Fokker-Planck solution of Eq.~(\ref{eq|fokkersol_barrier_full}) for a mass-dependent threshold $\Delta\delta_c(\sigma,t)$ modulating the noise, as given by Eqs.~(\ref{eq|barrier}) and (\ref{eq|barrauxvar}); the same shape parameters used in paper I to reproduce the unconditional halo mass function have been retained. Dashed line shows the corresponding result for the excursion set framework with a moving barrier. Dotted line refers to the EPS theory, i.e. a standard spherical collapse threshold $\Delta\delta_c(t)$ independent of mass. All the analytic results have been computed with the same cosmological parameters of the \emph{Millennium} $N$-body simulations for fair comparison.}\label{fig|condmultfunc}
\end{figure}

In Fig.~\ref{fig|condmultfunc} we illustrate the conditional multiplicity function $\nu_{12}\, \bar{\mathcal{P}}(\nu_{12})$ as a function of the scaled, self-similar variable $\nu_{12}$. The expectation from our stochastic theory based on the Fokker-Planck equation is compared to the outcomes from the \emph{Millennium} $N$-body simulations for FoF-identified halos by Cole et al. (2008). We find a reasonable agreement, and stress that this is obtained with the same mass-dependent threshold (i.e., the same parameters $q$, $\beta$, $\gamma$) used in paper I to reproduce the halo mass function. As a reference, in Fig.~\ref{fig|condmultfunc} we also illustrate the results expected from the excursion set theory for a spherical collapse (EPS) and for a mass-dependent collapse threshold (moving barrier). As it is well known, the former exceeds the $N$-body results at the low-mass end, and falls short at the high mass end; the latter performs quite well in the low/intermediate mass range, but exceeds somewhat at the high mass end.

In Fig.~\ref{fig|condmassfunc} we show the mass fraction in progenitor haloes $f(\log M_1|M_2)\equiv \mathcal{P}(M_1,t_1|M_2,t_2)$ $\times M_1\,\ln(10)$ as a function of the progenitor to descendant mass ratio $M_1/M_2$, for different descendant masses $M_2\sim 10^{12}$, $3\times 10^{13}$,  $10^{15}\, M_\odot$ at $z_2\approx 0$ and different progenitor redshifts $z_1\sim 0.5$, $1$, $2$, $4$. As above, we find an overall good agreement between the expectations from our stochastic theory based on the Fokker-Planck equation and the outcomes by the \emph{Millennium} $N$-body simulations by Cole et al. (2008); note that in the latter the merger trees have been self-consistently constructed by linking FoF groups which contain one or more subhalos between successive time-steps (more details can be found in Springel et al. 2005).

Two remarks are in order here. First, all the analytic results have been computed with the same cosmological parameters of the \emph{Millennium} simulations for fair comparison. Second, for self-consistency with the comparison pursued in paper I for the unconditional mass function, we confront the stationary solutions of Eq.~(\ref{eq|fokker_barrier}) from our stochastic theory to the outcomes of simulations where halos are identified via FoF-based algorithms, since both tend to produce closely universal functions in terms of the scaled, self-similar variable $\nu_{12}$. As mentioned in paper I, more general (yet not analytic) solutions from our stochastic theory constitute transitional states that, though converging to the stationary one in the long-time limit, could be possibly related to deviation of the mass function from the self-similar shape; a detailed comparison of these with the outcomes of numerical simulations based on spherical overdensity algorithm for halos identification would be welcome but beyond our scope here.

\section{Large-scale halo bias}\label{sec|bias}

Given the conditional distribution derived in the previous Section, it is straightforward to compute the large-scale halo bias, which is routinely exploited to evaluate the correlation between halo abundances and the surrounding environment. The (Eulerian) halo bias can be written as (see Mo \& White 1996; Sheth \& Tormen 1999)
\begin{equation}\label{eq|basic_bias}
b(M,t) = 1+\cfrac{1}{\delta_{c0}}\,\left[\cfrac{N(M,t|M_0,t_0)}{N(M,t)\, V}-1\right]~,
\end{equation}
in terms of the ratio between the number of halos at time $t$ (density contrast $\delta_c$) that will end up at time $t_0$ (density contrast $\delta_{c0}$) in an environment with volume $V$ and mass $M_0\equiv\bar\rho\, V$.

Recalling from paper I (see also Appendix A) the expression of the halo mass function
$N(M,t)=(\bar\rho/M)\,$ $\bar{\mathcal{P}}_{\rm MF}(\nu)\,|{\rm d}\nu/{\rm d}M|$ and using Eq.~(\ref{eq|prog_mf}), one can recast Eq.~(\ref{eq|basic_bias}) in the form
\begin{widetext}
\begin{equation}\label{eq|full_bias}
b(M,z) = 1+\cfrac{1}{\delta_{c0}}\,\left[\cfrac{\nu_{\rm CMF}\,\bar{\mathcal{P}}_{\rm CMF}(\nu_{\rm CMF})}{\nu\,\bar{\mathcal{P}}_{\rm MF}(\nu)}-1\right]\simeq 1-\cfrac{1}{\delta_{c}}\,\left[1+\cfrac{\nu\,\bar{\mathcal{P}}_{\rm MF}'(\nu)}{\bar{\mathcal{P}}_{\rm MF}(\nu)}\right]~,
\end{equation}
\end{widetext}
where $\nu_{\rm CMF}\equiv (\delta_c-\delta_{c0})/\sqrt{\sigma^2(M)-\sigma^2(M_0)}$ and $\nu\equiv \delta_c(t)/\sigma(M)$. The last equality approximately holds on large-scales where one can assume $\delta_{c0}\ll \delta_c$, $\sigma(M_0)\ll \sigma(M)$ and $\nu_0\equiv \delta_{c0}/\sigma(M_0)\gg 1$, so that at leading order $\nu_{\rm CMF}\simeq \nu-(\delta_{c0}/\delta_c)\, \nu$ and $\bar{\mathcal{P}}_{\rm CMF}(\nu_{\rm CMF})\simeq \bar{\mathcal{P}}_{\rm MF}(\nu)-(\delta_{c0}/\delta_c)\, \nu\, \bar{\mathcal{P}}_{\rm MF}'(\nu)$, where the prime means differentiation with respect to $\nu$. All in all, using the explicit expression of the unconditional probability Eq.~(\ref{eq|fokkersol_barrier_uncond}) we obtain
\begin{widetext}
\begin{empheq}[box=\widewidefbox]{align}\label{eq|largescale_bias}
b(M,z) \simeq 1+\cfrac{1}{\delta_c}\,\left\{-1+\cfrac{2\beta\gamma}{\beta+(\sqrt{q}\nu)^{2\gamma}}
+(\sqrt{q}\nu)^2\,\left[1+\cfrac{\beta}{(\sqrt{q}\nu)^{2\gamma}}\right]^2\right\}~.
\end{empheq}
\end{widetext}

\begin{figure}
\centering
\includegraphics[width=\textwidth]{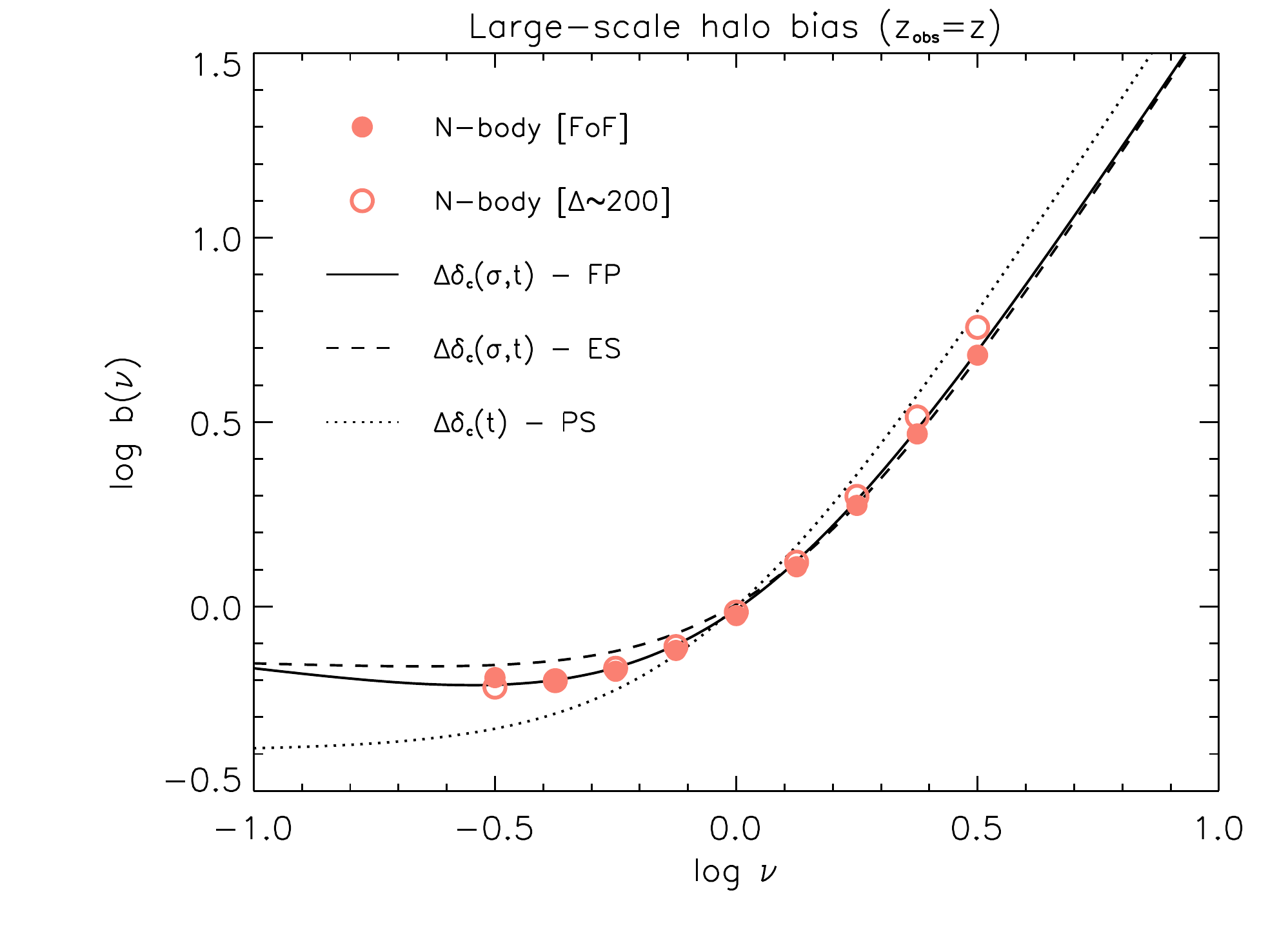}
\caption{The large-scale halo bias as a function of the self-similar, scaled variable $\nu\equiv \delta_c/\sigma$. Dots illustrate the outcome from $N$-body simulations: filled symbols refer to Tinker et al. (2005) where halos are identified with a FoF algorithm, empty symbols refer to Tinker et al. (2010) where halos are identified with a spherical overdensity algorithm with nonlinear threshold $\Delta=200$. Linestyles as in previous Figures.}\label{fig|bias}
\end{figure}

In Fig.~\ref{fig|bias} we illustrate the large-scale halo bias as a function of the variable $\nu\equiv \delta_c/\sigma$, and compare it with the outcomes from $N$-body simulations by Tinker et al. (2005) for halos identified by a FoF algorithm, and by Tinker et al. (2010) for halos identified by a spherical overdensity algorithm with non-linear threshold $\Delta=200$; we find a remarkable agreement with both determinations. In comparison, the EPS outcome for spherical collapse exceeds the simulation results for large $\nu$ (high masses and/or early times) and falls short for small $\nu$. The excursion set with moving barrier performs quite well for large $\nu$ but exceeds somewhat the $N-$body outcomes for $\nu\lesssim 1$.

\bigskip

\section{Summary and outlook}\label{sec|summary}

We have generalized the stochastic theory of hierarchical clustering presented in paper I by Lapi \& Danese (2020), so as to derive the (conditional) halo progenitor mass function and the related large-scale bias.
In Sect.~\ref{sec|stochastic} we have presented a stochastic differential equation that describes fluctuations in the mass growth of progenitor halos of given descendant mass and redshift, as driven by a multiplicative Gaussian white noise involving the spherical collapse thresholds and the mass variances of the progenitor and descendant halos. By numerically integrating the stochastic equation we have demonstrated that, as cosmic time passes, the noise yields an average drift of the progenitors toward larger masses, that quantitatively renders the expectation from the standard extended Press \& Schechter (EPS) theory. In Sect.~\ref{sec|fokker} we have solved the Fokker-Planck equation associated to the stochastic dynamics, and obtain as an exact solution the EPS progenitor mass function. We have also pointed out that the solution is stationary when written in terms of a convenient, scaled variable. In Sect.~\ref{sec|driftdiff} we have highlighted how the formalism based on the Fokker-Planck equation allows to define unambiguously the total derivative of the (conditional and unconditional) mass function in terms of halo direct collapse vs. merging rates, naturally solving an otherwise tricky issue in the excursion set formalism.

In Sect.~\ref{sec|barrier} we have introduced a modification of the stochastic equation in terms of a mass-dependent collapse threshold modulating the noise. We have solved analytically the associated Fokker-Planck equation for the progenitor mass function. The latter is found to be in excellent agreement with the outcome of $N-$body simulations. Even more remarkably, all that is achieved with the same threshold used in paper I to reproduce the halo mass function. In Sect.~\ref{sec|bias} we have exploited the progenitor mass function and the halo mass function to compute the large-scale halo bias, finding it in pleasing agreement with the $N-$body outcomes.

All in all, in the present paper we have illustrated that the stochastic theory of hierarchical clustering introduced in paper I can describe effectively not only the halos' abundance, but also their progenitor distribution and their correlation with the large-scale environment across cosmic times. Future developments will be focused on applying our stochastic theory to derive the statistics of cosmic voids, to build up refined merger-tree algorithms, and to investigate the distribution and clustering of sub-halos. It would be also worth exploring whether the same stochastic approach can be usefully applied to describe other phenomena in the modern astrophysical and cosmological landscape, such as galaxy formation or dark energy phenomenology.

\acknowledgments

We acknowledge the anonymous referee for a constructive report.
We warmly thank C. Baccigalupi and T. Ronconi for helpful comments and critical reading. This work has been partially supported by PRIN MIUR 2017 prot. 20173ML3WW 002, `Opening the ALMA window on the cosmic evolution of gas, stars and supermassive black holes'. A.L. acknowledges the MIUR grant `Finanziamento annuale individuale attivit\'a base di ricerca' and the EU H2020-MSCA-ITN-2019 Project 860744 `BiD4BEST: Big Data applications for Black hole Evolution STudies'.

\appendix

\section{Summary and a further test from paper I}

In this Appendix we recap the assumptions and results from paper I (Lapi \& Danese 2020), and provide a further test concerning the unconditional halo mass function in warm dark matter (WDM) cosmologies.

According to the standard cosmological framework, DM halos are thought to originate by the collapse of patches from an initial, closely Gaussian perturbation field. However, as demonstrated by many extensive $N-$body simulations (e.g., see textbook by Mo et al. 2010 and references therein), the detailed evolution of perturbations and proto-halo patches ultimately depends on a variety of effects. First, the role of initial conditions is crucial, in that a perturbation is more prone to collapse if it resided within a sufficiently overdense region of the initial density field. This was actually the idea at the basis of pioneering estimates for the halo abundance (see Press \& Schechter 1976), subsequently refined in terms of the excursion set approach (see Bond et al. 1991; Lacey \& Cole 1993; Mo \& White 1996) to avoid the double counting of overdense regions overlapped with, or embedded within larger collapsing ones (the so called cloud-in-cloud issue). In addition, the shape of proto-halo patches may be also relevant, in that they tend to be ellipsoidal in shape, and this may influence the collapse efficiency and timescale at different mass scales (see Sheth \& Tormen 1999, 2002). Moreover, collapse locations may be special points in the initial perturbation field (see Bardeen et al. 1986; Sheth et al. 2001; Dalal et al. 2008; Ludlow \& Porciani 2011; Musso \& Sheth 2012; Paranjape \& Sheth 2012; Hahn \& Paranjape 2014; Lapi \& Danese 2014), such as peaks in density or in energy (see Musso \& Sheth 2019). Finally, other nonlinear effects may influence the collapse of perturbations, such as the local environment, tidal forces, angular momentum, velocity fields, clumpiness and dynamical friction, etc.

From the above it should appear evident that the collapse and evolution of DM (proto)halos constitutes an inherently stochastic process, originated not only by a degree of randomness in the initial conditions, but also by the complexity of deterministic processes affecting the ensuing collapse. As a consequence, the fine details of the evolution for individual proto-halos at different spatial locations and cosmic times are, for all practical purposes, difficult to be followed and/or modeled ab-initio in (semi-)analytic terms. Nevertheless, in paper I we have submitted that, if one is interested mainly in the statistical properties of the halo population as a whole, an effective description could be conveniently obtained in terms of stochastic differential equations with appropriate noise.

The situation is somewhat analogous to the classic description of Brownian motion: a microscopic particle immersed in a fluid continuously undergoes collisions with the fluid molecules; the resulting motion $x_t$, despite being deterministic, appears to be random at the macroscopic level, especially to an external observer who has no access to the exact positions and velocities of the (innumerable) fluid molecules and to the initial conditions of the particle. In the way of a statistical description, the problem is effectively treated via a stochastic differential equation $\dot x_t\sim \kappa\, \eta(t)$, in terms of a diffusion constant $\kappa$ and of a fluctuating white noise $\eta(t)$, that allows to implicitly account for the complex microscopic dynamics of the system. Note that often the system's state influences the intensity of the driving noise, like when the Brownian fluctuations of a microscopic particle near a wall are reduced by hydrodynamic interactions, so that the noise becomes 'multiplicative' in terms of a non-uniform diffusion coefficient $\kappa=\kappa(x_t)$. Similar stochastic models with multiplicative noise have been employed to describe a wide range of physical phenomena, from Brownian motion in inhomogeneous media or in close approach to physical barriers, to thermal fluctuations in electronic circuits, to the evolution of stock prices, to computer science, to the heterogeneous response of biological systems and randomness in gene expression (e.g., Risken 1996; Reed \& Jorgensen 2003; Mitzenmacher 2004; Paul \& Baschnagel 2013).

Our proposal in paper I has been to adopt a similar description to capture the essence of the proto-halo collapse via the following nonlinear stochastic differential equation
\begin{equation}\label{eq|app_basic_sigma}
\cfrac{\rm d}{{\rm d}t}\ln\sigma^{-1}(M_t) = \cfrac{\sigma(M_t)}{\delta_c(t)}\, \left|\cfrac{\dot D}{D}\right|^{1/2}\, \eta(t)~,
\end{equation}
%or equivalently in terms of mass
%\begin{equation}\label{eq|app_basic_mass}
%\cfrac{\rm d}{{\rm d}t}\, M = \cfrac{\sigma^2}{|{\rm d}\sigma/{\rm d}M|}\, \cfrac{1}{\delta_c}\, \left|\cfrac{\dot D}{D}\right|^{1/2}\, \eta(t)~;
%\end{equation}
where $t$ is the cosmic time (corresponding to redshift $z$), $\eta(t)$ is a Gaussian white noise with ensemble-averaged properties $\langle\eta\rangle=0$ and $\langle\eta(t)\eta(t')\rangle=2\,\delta_D(t-t')$ (the factor $2$ is only a convention and clearly it could be reabsorbed into the multiplicative term), $\delta_c(\sigma,t)$ is the critical threshold for collapse (possibly dependent on scale, see below), $D(t)$ is the linear growth factor of perturbations, and $\sigma(M)$ is the mass variance filtered on the mass scale $M$ defined in Eq.~(\ref{eq|variance}).
The rationale naively followed to invent the above Eq.~(\ref{eq|app_basic_sigma}) is simple. On the left hand side it appears the time derivative of an adimensional function of the mass that incorporates the power spectrum and the filtering scale; in choosing $\ln\sigma$ we have been inspired by a number of $N$-body simulations (e.g., Zhao et al. 2009), that suggest the mass growth of halos to be easily described in terms of such a quantity. On the right hand side it appears a stochastic driving $\eta(t)$, that for dimensional consistency must be multiplied by the (inverse) square root of a characteristic timescale. Since our aim here is to describe the growth of DM perturbations using quantities related to the linear regime, we find it natural to choose the timescale $|\dot D(t)/D(t)|$ that effectively describes the linear growth of perturbations under gravity in a given cosmological background.

We now discuss the meaning and relevance of the various terms in Eq.~(\ref{eq|app_basic_sigma}). The mass variance $\sigma^2(M)\propto \int{\rm d}^3k\,P(k)\, \tilde{W}_M^2(k)$ incorporates the initial conditions in terms of the power spectrum of perturbations $P(k)$. We stress that in the present theory the relation $\sigma(M)$ between $\sigma$ and $M$ is purely deterministic;
however, both $M_t$ and $\sigma(M_t)$ are to be considered stochastic variables that fluctuate over cosmic time $t$ under the influence of the noise. This is a different perspective with respect to the standard excursion set formalism, where the overdensity field $\delta(\sigma)$ executes a random walk as a function of the mass variance $\sigma$, which in turn plays the role of a pseudo-time variable. We further note that in the excursion set approach the choice of the filter function $\tilde{W}_M^2$ in the definition of $\sigma(M)$ has a crucial impact, since the random trajectories $\delta(\sigma)$ are Markovian only when a $k-$sharp filter in Fourier space is adopted; in the present theory, assuming a different filter function (e.g., Gaussian or top-hat in real space) changes only the deterministic relation $\sigma(M)$ but has otherwise no effect on the Markovianity (but see paper I for a non-Markovian generalization) of the stochastic processes $M_t$ and $\sigma(M_t)$. From the purely technical point of view, our approach has some degree of similarity to the formulation of the excursion sets in terms of stochastic differential equations and path integral representations by Maggiore \& Riotto (2010a,b); however, in our framework  the stochastic behavior is not associated to the trajectories in the abstract $\delta(\sigma)$ space but to the physical variables $M_t$ or $\sigma(M_t)$; further key differences are also evident from the discussion below.

The fluctuating Gaussian white noise $\eta(t)$ is meant to implicitly describe, at a macroscopic level, both the stochasticity in the initial conditions and all the complex physical phenomena associated to nonlinearities in the gravitational evolution. In fact, the adopted noise is `multiplicative', in that its strength depends on the state of the system as expressed by the ratio $\sigma/\delta_c$. Patches with $\sigma\gtrsim \delta_c$ tend to change their mass more abruptly, while the evolution is slower for $\sigma\lesssim \delta_c$; this renders, in our stochastic theory, the naive expectation that overdense regions in the initial perturbation field are more prone to collapse. Nevertheless, there is a crucial difference with the excursion set theory. In the latter, to avoid double counting of overlapping patches, only the first crossings of the walk $\delta(\sigma)$ with the barrier at $\delta_c$ must be considered; this implies that, for the relevant trajectories, the mass of proto-halos can only increase.
In our approach the fluctuating noise can induce both positive and negative variations of the mass $M_t$ within the filtered region (with the obvious constraint $M_t>0$): positive variations of $M_t$ can be reasonably related to mergers among collapsing proto-halos, or mass accretion from the field; negative variations can be interpreted in terms of mass loss due to gravitational interactions, tidal forces, stripping, and fragmentation with overlapping or embedding proto-halo patches. On the other hand, the multiplicative nature of the noise induces, on statistical grounds, an average drift toward larger masses: as $\eta(t)$ fluctuates, also the random variable $\sigma(M_t)$ and hence the multiplicative factor $\sigma/\delta_c$ on the r.h.s. of Eq.~(\ref{eq|app_basic_sigma}) varies, and therefore $\langle\sigma\, \eta/\delta_c\rangle$ is not null even if $\langle\eta\rangle$ is. It turns out, as shown in paper I by the numerical integration of Eq.~(\ref{eq|app_basic_sigma}), that this noise-induced drift actually makes $\sigma(M_t)$ to copy the decrease of $\delta_c(t)$ with time, so that given the inverse convex shape of the deterministic function $\sigma(M)$ a net average increase in mass $M_t$ is enforced; this is of the correct amount to avoid the cloud-in-cloud issue and statistically recover, without the need to restrict to first crossing distributions, the excursion set result on halo abundance (at least when the mass-independent $\delta_c$ from spherical collapse is employed). In other words, the cloud-in-cloud issue is inherently treated and solved via the multiplicative noise structure of Eq.~(\ref{eq|app_basic_sigma}).

The collapse threshold $\delta_c(\sigma,t)$ is meant to represent the level of overdensity required for a perturbation on given mass scale to undergo efficient collapse. In paper I we have adopted a dependence on the mass variance analogous to what is done in the excursion set
\begin{equation}\label{eq|app_barrier}
\delta_c(\sigma,t)\approx  \sqrt{q}\delta_c\,\left[1+\beta\, \left(\frac{\sigma}{\sqrt{q}\delta_c}\right)^{2\gamma}\right]~,
\end{equation}
in terms of the spherical collapse threshold $\delta_c(t)\approx 1.68\,D(0)/D(t)$ and of three shape parameters $(q,\beta,\gamma)$. In the excursion set approach $\delta_c(\sigma,t)$ constitutes a deterministic barrier (but see Maggiore \& Riotto 2010b for generalizations) that the random walk $\delta(\sigma)$ must hit to enforce collapse; its mass dependence is ascribed, though a bit naively, to the fact that perturbations may be ellipsoidal in shape. In our stochastic framework, $\delta_c(\sigma,t)$ is just a quantity modulating the noise, and as such is meant to describe a bunch of effects not only limited to the shape of proto-halo patches, but also to other complex phenomena that can influence their collapse, like tidal torques and angular momentum, cosmological constant, dynamical friction, and possibly the speciality of collapse locations. In this vein, in paper I we have set the values of the parameters $(q,\beta,\gamma)$ in order to reproduce the unconditional halo mass function measured in $N-$body simulations.

Specifically, the probability density function $\mathcal{P}(M,t)$ of finding the system in a state with mass $M$ at cosmic time $t$ can be evaluated by writing down the Fokker-Planck equation associated to the stochastic Eq.~(\ref{eq|app_basic_sigma}), with the collapse threshold given by Eq.~(\ref{eq|app_barrier}); in terms of the self-similar, scaled variable $\nu\equiv \delta_c(t)/\sigma(M)$, this can be put in the compact form
\begin{equation}\label{eq|app_fokker}
\partial_t\,\mathcal{P}(\nu,t)=\left|\cfrac{\dot D}{D}\right|\,
\partial_\nu\,\left\{\nu\,\mathcal{P}(\nu,t)+\cfrac{1}{2\mathcal{P}(\nu,t)}\,\partial_\nu\,
\left[\cfrac{\mathcal{P}(\nu,t)}{\sqrt{q}\,\left[1+\beta/(\sqrt{q}\nu)^{2\gamma}\right]}\right]^2\right\}~.
\end{equation}
As discussed in paper I, the halo mass function $N(M,t)$ can be related to the stationary solutions $\bar{\mathcal{P}}(\nu)$ of the above equation; the result turns out to be
\begin{equation}\label{eq|app_massfunc}
N(M,t)= \cfrac{\bar\rho}{M\,\sigma}\, \left|\cfrac{{\rm d}\sigma}{{\rm d}M}\right|\,\nu\,\bar{\mathcal{P}}(\nu)~,
\end{equation}
where
\begin{equation}\label{eq|app_multfunc}
\bar{\mathcal{P}}(\nu)= \mathcal{A}\, \sqrt{q}\,\left[1+\cfrac{\beta}{(\sqrt{q}\,\nu)^{2\gamma}}\right]\, \exp\left\{-q\,\cfrac{\nu^2}{2}\,\left[1+\cfrac{2\,\beta}{1-\gamma}\,\cfrac{1}{(\sqrt{q}\,\nu)^{2\gamma}}
+\cfrac{\beta^2}{1-2\,\gamma}\,\cfrac{1}{(\sqrt{q}\,\nu)^{4\gamma}}\right]\right\}~,
\end{equation}
and $\mathcal{A}$ is a constant determined by the normalization condition $\int_0^\infty{\rm d}\nu\, \bar{\mathcal{P}}(\nu)=1$. When posing $q=1$ and $\beta=0$ corresponding to the mass-independent, spherical collapse threshold $\delta_c(t)$, we exactly
recover the Press \& Schechter (1976) mass function $\bar{\mathcal{P}}(\nu)=\sqrt{2/\pi}\,e^{-\nu^2/2}$. On the other hand, comparison with the outcomes of $N-$body simulations (see Sheth \& Tormen 1999; Bhattacharya et al. 2011; Watson et al. 2013) yields the optimal parameter values $q\approx 0.65$, $\beta\approx 0.15$, $\gamma\approx 0.35$.
Remarkably, the asymptotic behavior for $\nu>>1$, corresponding to large masses and/or early cosmic times, is seen to produce a shape akin to the empirical fit of $N$-body simulations adopted since Sheth \& Tormen (1999); however, for finite $\nu$ the terms in the exponential are important and must be taken into account.

\subsection{Halo mass function in warm dark matter cosmologies}

We now provide a further test of our stochastic theory, concerning the halo mass function in warm darm matter (WDM) cosmologies. In fact, if our framework has captured the macroscopic essence of the proto-halo evolution and collapse, it should be able to describe the halo mass function for cosmologies different from the standard CDM, and in particular for WDM models, without any adjustment of the parameters entering the expression of the halo mass function (e.g., Hahn \& Paranjape 2014). To wit, the halo mass function should be still expressed by Eqs. (\ref{eq|app_massfunc}) and (\ref{eq|app_multfunc}) with the same fitted parameter values $(q,\beta,\gamma)$ given above, and the only modification with respect to CDM would be in the deterministic relation $\sigma(M)$ that depends explicitly on the power spectrum of perturbations.

Specifically, we employ a WDM power spectrum (Bode et al. 2001; Viel et al. 2005) given by
\begin{equation}
P_{\rm WDM}(k) = P_{\rm CDM}(k)\, \left[1+(\alpha\, k)^{2\mu}\right]^{-10/\mu}
\end{equation}
with $P_{\rm CDM}(k)$ the standard CDM spectrum (e.g., Bardeen et al. 1986), $\mu\approx 1.12$ and
\begin{equation}
\alpha\approx 0.049\, \left(\cfrac{m_{\rm WDM}}{\rm keV}\right)^{-1.11}\, \left(\cfrac{\Omega_{\rm WDM}}{0.25}\right)^{0.11}\, \left(\cfrac{h}{0.7}\right)^{1.22}\, h^{-1}~~{\rm Mpc}~,
\end{equation}
in terms of the WDM mass $m_{\rm WDM}$ for a thermally produced particle;
WDM free-streaming effects introduce a characteristic mass scale of suppression $M_{\rm WDM}=(4\pi/3)\, \bar\rho\left[\pi\alpha\,(2^{\mu/5}-1)^{-1/2\mu}\right]^3$ in the power spectrum. We consider WDM masses $m_{\rm WDM}\approx 0.25$, $0.5$, and $1$ keV, corresponding to suppression scales $M_{\rm WDM}\approx 10^{10}$, $10^{11}$ and $10^{12}\, M_\odot$. We note that such particle masses are already ruled out by observations of the Lyman-$\alpha$ forest and high$-z$ galaxy statistics (see Viel et al. 2013; Lapi \& Danese 2015), and are considered here just for the sake of testing.
From the expression of the power spectrum, we compute the mass variance $\sigma^2(M)\propto \int{\rm d}^3k\,P_{\rm WDM}(k)\, \tilde{W}_M^2(k)$ by choosing a $k-$sharp filter in Fourier space $\tilde W_M(k)=\theta_{\rm H}(k_M-k)$ with standard mass assignment $k_M=(6\pi^2\,\bar \rho/M)^{1/3}=(9\pi/2)^{1/3}/R_M$ and $R_M=(3\,M/4\pi\bar\rho)^{1/3}$; note that we do not find necessary to change the multiplicative factor in the relation $k_M\propto R_M^{-1}$ as in Benson et al. (2013) and Schneider et al. (2013). As discussed by these authors, the choice of the $k-$sharp filter is motivated empirically, since it originates a mass function which gets correctly suppressed below the free-streaming length of the WDM particles; possibly a deeper reason lays in that the real-space non-locality inherent in the filter, that may somehow capture well the properties of the initial density environment near small-mass WDM peaks (see Hahn \& Paranjape 2014). For completeness, we will also show results for a real-space top-hat window $\tilde{W}_M(k)=3\,(kR_M)^{-3}\, [\sin(kR_M)-kR_M\,\cos(kR_M)]$; a Gaussian filter produces indistinguishable outcomes.

\begin{figure}
\figurenum{A1}
\centering
\includegraphics[width=\textwidth]{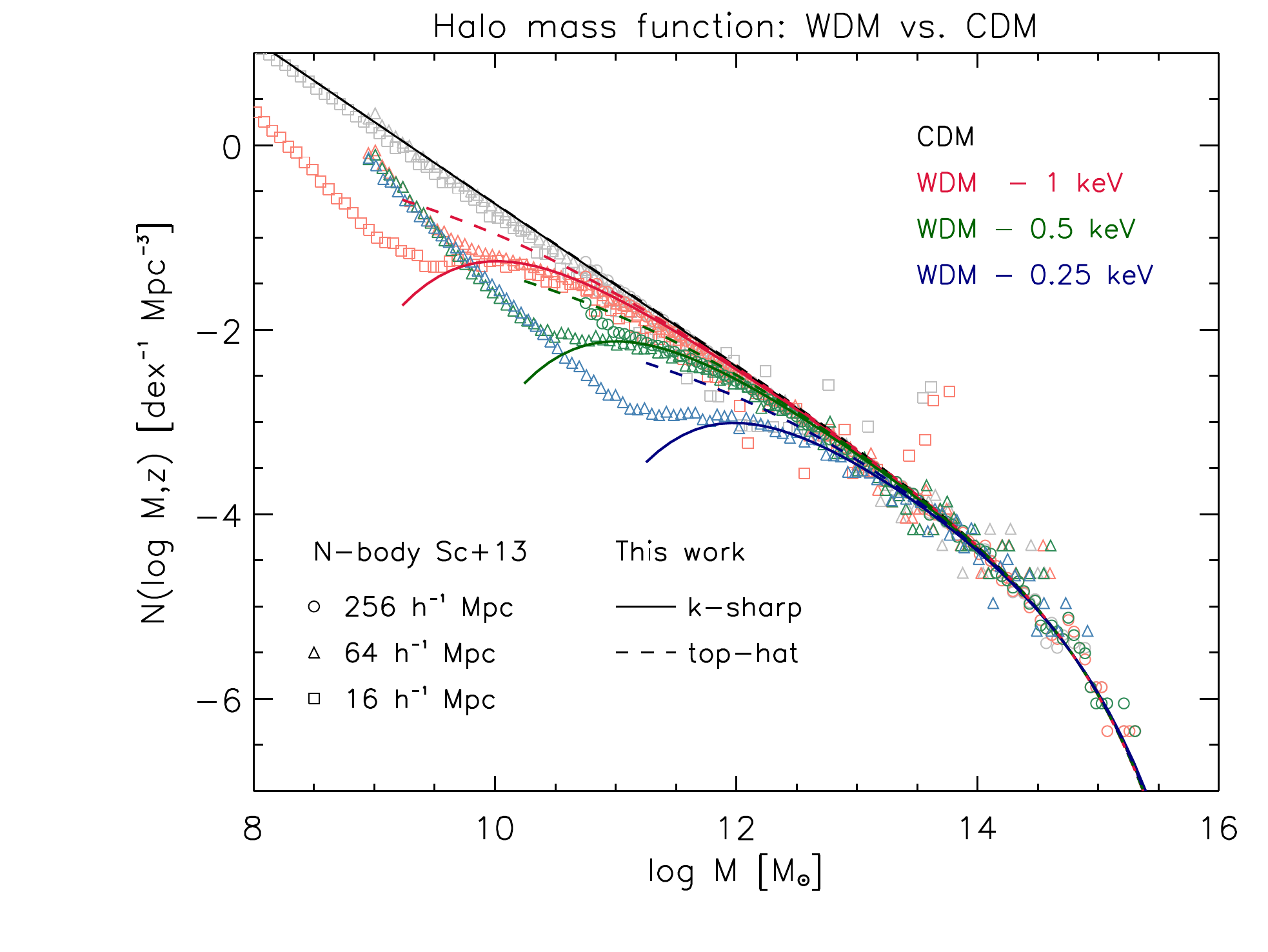}
\caption{The unconditional halo mass function at $z=0$ in WDM cosmologies. Black lines and symbols refer to standard CDM, red ones to WDM with particle mass $m_{\rm WDM}\approx 1$ keV, green ones to WDM with $m_{\rm WDM}\approx 0.5$ keV and blue ones to WDM with $m_{\rm WDM}\approx 0.25$ keV. Symbols show the outcomes of the $N-$body simulations by Schneider et al. (2013) based on FoF halo identification, $1024^3$ particles and three different resolutions corresponding to box sizes $256\, h^{-1}$ Mpc (circles), $64\, h^{-1}$ Mpc (triangles), and $16\, h^{-1}$ Mpc (squares); the increase in the WDM simulations below the free-streaming mass scale is spuriously caused by artificial fragmentation. Solid and dashed lines illustrate the expectations of our stochastic theory based on the Fokker-Planck equation, when adopting a $k-$sharp filter (solid lines) or a top-hat/Gaussian filter (dashed lines; superimposed to solid line for the CDM case) in the computation of the deterministic relation $\sigma(M)$, see Appendix A for details.}\label{fig|WDM}
\end{figure}

The predictions on the WDM mass function from Eqs.~(\ref{eq|app_massfunc}) and (\ref{eq|app_multfunc}) at $z=0$ are illustrated in Fig.~\ref{fig|WDM} as solid ($k-$sharp filter) and dashed (real top-hat or Gaussian filter) lines, for three values of the WDM particle mass $m_{\rm WDM}=0.25$ (blue), $0.5$ (green) and $1$ (red) keV; the CDM mass function is also plotted for reference (black line). The WDM mass functions are found to deviate from the CDM one around the corresponding free-streaming mass scale $M_{\rm WDM}$, with the $k-$sharp filter producing the expected downturn. Our results for $k-$sharp filter well agree with the outcomes from the $N-$body simulations by Schneider et al. (2013). Two caveats are worth to be mentioned on the simulated WDM mass functions: (i) the steepening occurring much below the free-streaming mass scale is well known to be spuriously caused by artificial fragmentation (see also discussion by Hahn \& Paranjape 2014); (ii) the effects of late-time thermal velocities are neglected or approximately treated, so that the shape of the halo mass function around the free-streaming mass scale is still debated (see Benson et al. 2013). Given these uncertainties, we consider the result of the test conducted here quite satisfying.

\newpage

\newpage

\newpage

\newpage

\newpage

\end{document}